\documentclass[sigconf]{acmart}


\usepackage[inline]{enumitem}

 \usepackage{balance}

\begin{document}

\title{\texttt{DESERE}: The 1st Workshop on Decentralised Search and Recommendation}
%\subtitle{\url{https://rgm-cikm23.github.io}}
% \renewcommand{\shorttitle}{The 1st Workshop on Recommendation with Generative Models}

\author{Mohamed Ragab}
\email{ragab.mohamed@soton.ac.uk}
\affiliation{
\institution{University of Southampton}
\country{UK}
}

\author{Yury Savateev}
\email{y.savateev@soton.ac.uk}
\affiliation{%
  \institution{University of Southampton}
  \country{UK}
}
\authornote{Both first and second authors contributed equally to this research.}

\author{Wenjie Wang}
\email{wangwenjie@u.nus.edu}
\affiliation{%
  \institution{National University of Singapore}
  \country{Singapore}
}

\author{Reza Moosaei}
\email{r.moosaei@qmul.ac.uk}
\affiliation{%
  \institution{Queen Mary University of London}
  \country{UK}
}

\author{Thanassis Tiropanis}
\email{t.tiropanis@southampton.ac.uk}
\affiliation{
\institution{University of Southampton}
\country{UK}
}

\author{Alexandra Poulovassilis}
\email{a.poulovassilis@bbk.ac.uk}
\affiliation{%
  \institution{Birkbeck University of London}
  \country{UK}
}

\author{Adriane Chapman}
\email{adriane.chapman@soton.ac.uk}
\affiliation{
\institution{University of Southampton}
\country{UK}
}

\author{Helen Oliver}
\affiliation{%
  \institution{Birkbeck, University of London}
  \city{London}
  \country{UK}
}
\email{h.oliver@bbk.ac.uk}

\author{George Roussos}
\email{g.roussos@bbk.ac.uk}
\affiliation{%
  \institution{Birkbeck University of London}
  \country{UK}
}

% \author{Chua Tat Seng}
% \email{dcscts@nus.edu.sg}
% \affiliation{
% \institution{National University of Singapore}
% \country{Singapore}
% }

\renewcommand{\shortauthors}{Mohamed Ragab et al.}

\begin{abstract}
%Recent controversies over access and processing of personal data on the web have highlighted the significance of the sovereignty of individuals over their personal data and are leading to new paradigms for application development based on personal online datastores (pods), where individuals have complete control over which applications can gain access to their personal data and for what purpose. Emerging frameworks and ecosystems, such as SOLID, support the development of such decentralised applications which, when granted access by the individuals concerned, can access the data stored in pods to provide services to users in areas such as health and well-being, social networking, and collaborative authoring. However, this decentralisation presents significant performance challenges, which is critical to fulfilling the potential of such applications. 

The \texttt{DESERE} Workshop, our First Workshop on Decentralised Search and Recommendation, offers a platform for researchers to explore and share innovative ideas on decentralised web services, mainly focusing on three major topics:
\begin{enumerate*}[label=(\roman*)]
\item societal impact of decentralised systems: their effects on privacy, policy, and regulation; 
\item decentralising applications: algorithmic and performance challenges that arise from decentralisation; and 
\item infrastructure to support decentralised systems and services: peer-to-peer networks, routing, and performance evaluation tools. 
\end{enumerate*}

% The related research will improve the understanding and development of decentralized web services and contribute to new tasks and technologies in both academia and industry. 
% In the long run, we hope that this research direction will lead to new ways that personal data is stored and processed everywhere with new systems that support it. 
\end{abstract}

\maketitle

\section{Introduction}
Recent controversies over access and processing of personal data have highlighted the significance of the sovereignty of individuals over their personal data \cite{kahle2015locking} and are
leading to new paradigms for application development based on personal online datastores (pods) \cite{dedecker2022s}. 
Emerging frameworks and ecosystems, such as Solid \cite{sambra2016solid} and Dataswyft's\footnote{https://www.dataswyft.io/} Hub-of-All-Things (HAT), support
the development of such decentralised applications which, when granted access by the individuals concerned, can access the data stored in personal data stores (PDSes) to provide services to users in areas such as health and well-being, social networking, federated recommendation, and collaborative authoring \cite{sambra2016building}. However, this decentralisation presents significant performance challenges, which are of critical importance in fulfilling the potential of such application~\cite{Ragab2023}.

\begin{table*}[t!]
\centering
\setlength{\abovecaptionskip}{0.1cm}
\setlength{\belowcaptionskip}{0cm}
\caption{Program schedule.}
\label{tab:program}
\begin{tabular}{p{0.7\textwidth}cl}
\hline
\textbf{Event} & & \textbf{Time} \\
\hline
Opening Remarks& DESERE Organisers & 13:30--13:45 \\
Empowering Secure Search on the Decentralised Web & ESPRESSO project team & 13:45--14:15 \\
Structured Query Execution across Decentralized Environments & Dr. Ruben Taelman & 14:15--14:45 \\
Tea/Coffee Break & -- & 14:45--15:15 \\
Paper Session &  & 15:15--15:45 \\
\hspace{1em}• An Investigation into the Feasibility of Performing Federated Learning on Solid Servers & Nayil Arana & 15:15--15:30 \\
\hspace{1em}• LLM-based Federated Recommendation & 	Jujia Zhao & 15:30--15:45 \\
Keynote Talk on Trustworthy Federated Learning & Prof. Han Yu & 15:45--16:15 \\
Panel discussion with QA: Challenges \& future of decentralised search and recommendation & Panelists TBD & 16:15--16:45 \\
Closing Remarks & DESERE Organisers & 16:45--17:00\\
\hline
\end{tabular}
\end{table*}

The DESERE workshop focuses on the important and underdeveloped area of \textbf{decentralised services for search and recommendation} ~\cite{tiropanis2021search, Taelman2020, crestani2013distributed,moawad2019minaret,zhao2024llm}. 
%The workshop aims to discuss the challenges and opportunities arising from adopting decentralised systems and developing the necessary services. 
%
The primary goal of the DESERE workshop is to advance innovative studies within the realm of decentralised web services, with a particular emphasis on three main areas: 
\begin{itemize}
    \item Firstly, the DESERE Workshop aims to underscore the significance of examining how decentralising personal data storage can impact society, including enhancing data privacy, sovereignty, and governance. 
    \item Secondly, the DESERE Workshop seeks to motivate researchers to address the algorithmic and performance hurdles linked with decentralisation, investigating the potential benefits and costs of decentralizing applications. 
    \item Last but not least, DESERE aims to discuss the necessary infrastructure for decentralization, such as overlay networks, peer-to-peer systems, and innovative approaches to routing and information retrieval, alongside the creation of new benchmarks and standards for assessing the effectiveness of these systems.
\end{itemize}

DESERE aims to provide a pivotal platform for researchers to share their innovative ideas, approaches, and breakthroughs; while fostering an environment that encourages interdisciplinary collaboration and the exploration of novel applications.

The workshop will inspire researchers to explore new frontiers in decentralised systems, empowering them with services such as search and recommendation. 
This research direction has the potential to revolutionise the traditional data storage paradigm and lead to the development of next-generation decentralised systems. 
Different applications can be decentralised with these new techniques, from social media to machine learning. This confluence of related studies and technologies will bring novel features to industry products and contribute to new academic research advances. 

\section{Workshop Program}
The workshop will be held over \textbf{half a day} (PM). Each accepted paper will be given around 10 minutes for presentation and 5 minutes for QAs. In addition, we have invited two senior researchers in the field to give 30-minute keynote talks (Dr. Ruben Taelman, postdoctoral researcher at IDLab (https://www.rubensworks.net/), and Prof. Han Yu, (https://personal.ntu.edu.sg/han.yu/). The draft workshop programme schedule is shown in Table~\ref{tab:program}.

\section{Scope and topics}
\label{scope}

% The primary aim of this workshop is to foster innovative research in the field of decentralized web services, specifically focused on three key topics. First, we want to highlight the importance of exploring the societal implications of decentralization of personal data storage and the ways to improve data privacy, sovereignty, and regulation.
% Second, we want to encourage researchers to take on algorithmic and performance challenges associated with decentralization, and to explore what could be gained by decentralizing applications and at what cost. Last, the workshop welcomes innovations in the infrastructure necessary for decentralization --- overlay networks, peer-to-peer systems, and new ways of routing and information retrieval. This includes the development of new evaluation metrics and standards to measure the efficiency of such systems.

% The workshop will serve as a platform for researchers to contribute the latest breakthroughs in this very important field. 
We invited original submissions on decentralized systems, including but not limited to the following topics:
\begin{itemize}[leftmargin=*]
    % across various modalities, such as text, image, and video. 
    \item Algorithmic challenges in developing decentralized applications for search, information retrieval, recommendation, query processing, data federation, and machine learning. 
    \item Decentralized architectures for LLM-based applications.
    \item Trustworthy personal data store infrastructures.
    \item Security, privacy, and sovereignty for sensitive personal data, \textit{e.g.,} medical data. 
    \item Regulation required for supporting personal data storage ecosystems.
    \item Performance evaluation for decentralized systems and benchmarking frameworks~\cite{ragab2021bench}. 
    \item Routing, metadata, and query propagation in peer-to-peer systems.
    %\item Personal data storage systems.
    %\item Personal data storage reliability. 
\end{itemize}

\section{Call for papers}

    \subsection{Submission Information}
    % Participation and Selection Process
    % Describe the attendee participation and selection/review criteria and process
    
    % Prospective authors are invited to submit original research papers that address the related topics of interest. Submitted papers should not exceed 8 pages in length, including figures, tables, and references. Papers must be formatted according to the CIKM 2023 format and submission guidelines. 
    %Submission Guidelines:
    Submitted papers must follow the Web Conference guidelines for submissions (see \url{https://www2024.thewebconf.org/calls/short-papers/}).
    %Submitted papers (.pdf format) must use the template of ACM CIKM 2023. Please remember to add Concepts and Keywords. Submissions can be of varying length from 4 to 8 pages, plus additional pages for the reference pages. There is no distinction between long and short papers, but the authors may decide on the appropriate length of the paper. 
    All papers will undergo the same review process and review period. Paper submissions must conform to the ``double-blind'' review policy. All papers will be peer-reviewed by experts in the field. Acceptance will be based on relevance to the workshop, scientific novelty, and technical quality. Submission site: \url{https://easychair.org/conferences/overview?a=32385050}. 

 \vspace{-0.3cm}

    \subsection{Participation and Selection Process}
    The organizers will assign each submission to at least three program committee members without COI for review. After collecting the reviewers' comments, the organizers will convene a meeting to discuss the reviews of each submission and make the final decision. 

 \vspace{-0.3cm}
    
    \subsection{Tentative Timeline}
    \begin{itemize}
        %\item Submissions deadline: August 18, 2023 
        %\item Paper acceptance notification: September 15, 2023 
        %\item Workshop date: October 22, 2023
        \item Paper submission deadline: February 24, 2024
        \item Notification to authors: March 4, 2024
        \item Camera ready deadline: March 11, 2024
        \item Workshop date: May 13, 2024
    \end{itemize}
     Please note that this is a preliminary timeline. We are keen to align it with timelines shared by The Web Conference 2024 chairs. 
    
    % \subsection{Tentative Program Committee}
    
    % \noindent {Potential program committee members} are as follows:\\
    % \noindent \textbf{Tat-Seng Chua}, National University of Singapore\\
    % \noindent \textbf{Xiangnan He}, University of Science and Technology of China\\
    % \noindent \textbf{Barrie Kersbergen}, University of Amsterdam\\
    % \noindent \textbf{Chen Gao}, Tsinghua University\\
    % \noindent \textbf{Leigang Qu}, National University of Singapore\\

\section{Organisation}

\subsection{Workshop Chairs }
\begin{itemize}[leftmargin=*]

\item\textbf{Thanassis Tiropanis}, University of Southampton 

\item\textbf{George Roussos}, Birkbeck, University of London 

\item\textbf{See-Kiong Ng}, National University of Singapore 

\end{itemize}

\subsection{Organisation Committee}
\emph{Organisers} and \emph{Co-Chairs} of this workshop are as follows:
\begin{itemize}[leftmargin=*]

\item\textbf{Mohamed Ragab}, University of Southampton -- \textbf{Co-Chair}
\item\textbf{Yury Savateev}, University of Southampton -- \textbf{Co-Chair}
\item\textbf{Wenjie Wang}, National University of Singapore -- \textbf{Co-Chair}
\item\textbf{Reza Mooseai}, Queen Mary University of London 
\item\textbf{Helen Oliver}, Birkbeck, University of London 
\end{itemize}

\subsection{Program Committee}
\begin{itemize}[leftmargin=*]
    \item\textbf{Adriane Chapman}, University of Southampton
    \item\textbf{Radwa Elshawi}, University of Tartu 
    \item\textbf{Essam Mansour}, Concordia University 
    \item\textbf{Alexandra Poulovassilis}, Birkbeck, University of London
    \item\textbf{George Roussos}, Birkbeck, University of London
    \item\textbf{Ruben Taelman}, Ghent University
    \item\textbf{Thanassis Tiropanis}, University of Southampton
    \item\textbf{Wenjie Wang}, National University of Singapore     

\end{itemize}

\section{Organisers Bio}
    \begin{itemize}[leftmargin=*]

 \item 
    \noindent\textbf{Mohamed Ragab}\\
    \underline{Email}: ragab.mohamed@soton.ac.uk\\
    \underline{Affiliation}: University of Southampton\\
    \underline{Biography}:
    Mohamed Ragab works as a Research Fellow in the Web and Internet Science Group at the School of Electronics and Computer Science, Southampton University. At Southampton University, Ragab is a member of the ESPRESSO research project. Ragab obtained his Ph.D. in Computer Science from Tartu University, where his research focused on deciding the optimal configurations of Big Data platforms while processing large Knowledge Graphs~\cite{ragab2021bench,sakr2021future,ragab2020towards,moaawad2017fly}. His work resulted in several peer-to-peer reviewed publications in several prominent journals and conferences in the area of knowledge management, such as the ISWC, SWJ, CACM, DOLAP, and IEEE Big Data. He also was awarded the best Ph.D. symposium paper award at the 26th ADBIS conference in Torino, Italy~\cite{ragab2022towards}. Ragab’s research interests include Semantic Web, Large Knowledge Graphs processing, and Big Data analytics. His current research interests focus on decentralised web search and information retrieval.

    \item %\smallskip
    \noindent\textbf{Yury Savateev}\\
    \underline{Email}: y.savateev@soton.ac.uk\\
    \underline{Affiliation}: University of Southampton\\
    \underline{Biography}: Yury Savateev works on the ESPRESSO project at the University of Southampton (espressoproject.org) that aims at enabling efficient search over personal data stores, ensuring privacy and user sovereignty. His responsibilities include the development of usage scenarios to inform the system design, the development of appropriate indexing structures, and the setting up of benchmarking and testing environments.
    He obtained his Ph.D. in Mathematics from Moscow State University in 2009, and since then has worked in different areas, mainly in Mathematical Logic and Theoretical Computer Science. His research interests include automata theory, proof theory, and knowledge representation.

    \item \textbf{Reza Mooseai}\\
    \underline{Email}: r.moosaei@qmul.ac.uk\\
    \underline{Affiliation}: Queen Mary, University of London\\
    \underline{Biography}: Reza Moosaei is a Lecturer at Queen Mary University of London and holds honorary postdoctoral researcher positions with Southampton University and Birkbeck, University of London. He has over 20 years of experience in research and teaching. His primary research interests are in knowledge graphing (semantic web) and decentralised distributed search. Currently, he is engaged in research with the ESPRESSO project team, working on the development of decentralised distributed search across personal online data resources. 
    % Reza Moosaei holds the distinction of being the author of publications that have recognition and acknowledgment for their contributions to the field of knowledge graphs (semantic web). 
    
    \item 
    \noindent\textbf{Wenjie Wang}\\
    \underline{Email}: wangwenjie@u.nus.edu\\
    \underline{Affiliation}: National University of Singapore\\
    \underline{Biography}: Dr. Wenjie Wang is a research fellow at National University of Singapore (NUS). He received Ph.D. in Computer Science from NUS, supervised by Prof. Tat-Sent Chua. Dr. Wang was a winner of Google Ph.D. Fellowships. His research interests cover recommender systems, data mining, and causal inference. His first-author publications appear in top conferences and journals such as SIGIR, KDD, WWW, WSDM, and TOIS. His work has been selected into ACMMM 2019 Best Paper Final List. Moreover, Dr. Wang has served as the PC member and reviewer for the top conferences and journals including TPAMI, TOIS, TKDE, SIGIR, WWW, WSDM, and KDD. He has rich workshop and tutorial organisation experience at CIKM'23, SIGIR'23, and WWW'22. 

    \item \textbf{Thanassis Tiropanis}\\
    \underline{Email}: t.tiropanis@soton.ac.uk\\
    \underline{Affiliation}: University of Southampton\\
    \underline{Biography}:
   Thanassis is a professor at the University of Southampton, School of Electronics and Computer Science (ECS), Web and Internet Science Group. He is also affiliated with the Institute for Life Sciences and the Southampton Marine and Maritime Institute at the University. His expertise is in decentralised data architectures, distributed linked data infrastructures, semantic technologies with application in education, and the study of the evolution and impact of the Web and the Internet as socio-technical artefacts. He has over 25 years of research and teaching experience through posts at University College London (UCL), the Athens Information Technology (AIT) Institute, and the University of Southampton. He was a Visiting Associate Professor at the Department of Computer Science, National University of Singapore in 2017-2019. He is a fellow and a chartered IT professional with the BCS, a senior member of IEEE, a fellow of the Higher Education Academy in the UK, a member of the ACM, and a member of the Technical Chamber of Greece.

    \item 
    \noindent\textbf{Alexandra Poulovassilis}\\
    \underline{Email}: a.poulovassilis@bbk.ac.uk\\
    \underline{Affiliation}: Birkbeck, University of London\\
    \underline{Biography}:
    Alexandra is a Professor Emerita and College Fellow at Birkbeck, University of London. She worked for IBM Greece and held posts at University College London and King's College London before coming to Birkbeck in 1999. Her research interests are in data management, querying, analysis, integration, visualisation and personalisation, and she has held numerous research grants in these areas. As well as conducting fundamental research in computer science, much of her work involves forging interdisciplinary links across the sciences, social sciences and arts. At Birkbeck she has been Co-Director of the London Knowledge Lab (2003-2015), Founding Head of the Graduate Research School (2003-5), Founding Director of the Birkbeck Knowledge Lab (2016-2021), Assistant Dean and subsequently Deputy Dean for Research Enhancement in the School of Business, Economics \& Informatics (2009-21), and Head of Computer Science \& Information Systems (2003-6, 2009-10). She served on the Computer Science and Informatics sub-panel of the UK RAE 2008 and REF 2014 and is a Fellow of the British Computer Society.

    \item  \textbf{Adriane Chapman}\\
    \underline{Email}: adriane.chapman@soton.ac.uk\\
    \underline{Affiliation}: University of Southampton\\
    \underline{Biography}:
    Adriane is a Professor of Computer Science and Head of the Digital Health Research Group at the University of Southampton. Her research focuses on using data appropriately and effectively. This involves solving problems that span the areas of databases, dataset retrieval, provenance, consent, algorithmic accountability, fairness, and explanations. She has worked closely with the US Federal government, and influenced the Office of the National Coordinators (ONC) report on the usage of provenance within electronic health systems. She has advised the US Food and Drug Agency (FDA), the National Geospatial-Intelligence Agency (NGA), and the Department of Homeland Security (DHS) on data management problems.

    \item  \textbf{Helen Oliver}\\
    \underline{Email}: helen.oliver@bbk.ac.uk\\
    \underline{Affiliation}: Birkbeck, University of London\\
    \underline{Biography}:
    Helen is a Postdoctoral Research Fellow at the School of Computing and Mathematical Sciences at Birkbeck, University of London, working on the ESPRESSO research project; and a Visiting Academic at the University of Southampton. She holds a doctorate in Computer Science from the University of Cambridge, as well as degrees from the University of Oxford, including a Master's in Software Engineering with a focus on ontology design in OWL-RDF. She has 15 years' experience in research and 25 years' combined industry and academic experience in software development. Her interests can be found at the intersection of privacy; Web science; speculative and participatory design of wearable, pervasive and ubiquitous systems; security; and creative technology. She is a Member of the BCS; a Trustee of the Space Science, Engineering \& Environmental Foundation (https://ssef.org.uk/); and a member of the Coalition for Independent Technology Research.

    \item  \textbf{George Roussos}\\
    \underline{Email}: g.roussos@bbk.ac.uk\\
    \underline{Affiliation}: Birkbeck, University of London\\
    \underline{Biography}:
    George is the Professor of Pervasive Computing and Head of the School of Computing and Mathematical Sciences at Birkbeck College, University of London. He holds degrees from the Universities of Athens and Manchester and a doctorate from Imperial College London. Before joining Birkbeck he worked as the Research and Development Manager for a multinational technology corporation and the Internet Systems Security Officer for the Greek Ministry of Defence. His research interests include digital healthcare supported by mobile and wearable technology and privacy-preserving technologies for the decentralised web. He is a member of the ACM, the IEEE Computer Society, and a Data Protection Officer under IAPP.

    \end{itemize}

\bibliographystyle{ACM-Reference-Format}
\balance
\bibliography{main}


\begin{thebibliography}{15}

%%% the Web address.

\ifx \showCODEN    \undefined \def \showCODEN     #1{\unskip}     \fi
\ifx \showDOI      \undefined \def \showDOI       #1{#1}\fi
\ifx \showISBNx    \undefined \def \showISBNx     #1{\unskip}     \fi
\ifx \showISBNxiii \undefined \def \showISBNxiii  #1{\unskip}     \fi
\ifx \showISSN     \undefined \def \showISSN      #1{\unskip}     \fi
\ifx \showLCCN     \undefined \def \showLCCN      #1{\unskip}     \fi
\ifx \shownote     \undefined \def \shownote      #1{#1}          \fi
\ifx \showarticletitle \undefined \def \showarticletitle #1{#1}   \fi
\ifx \showURL      \undefined \def \showURL       {\relax}        \fi
% invisible to TeX
\providecommand\bibfield[2]{#2}
\providecommand\bibinfo[2]{#2}
\providecommand\natexlab[1]{#1}
\providecommand\showeprint[2][]{arXiv:#2}

\bibitem[Crestani and Markov(2013)]%
        {crestani2013distributed}
\bibfield{author}{\bibinfo{person}{Fabio Crestani} {and} \bibinfo{person}{Ilya Markov}.} \bibinfo{year}{2013}\natexlab{}.
\newblock \showarticletitle{Distributed information retrieval and applications}. In \bibinfo{booktitle}{\emph{Advances in Information Retrieval: 35th European Conference on IR Research, ECIR 2013, Moscow, Russia, March 24-27, 2013. Proceedings 35}}. Springer, \bibinfo{pages}{865--868}.
\newblock


\bibitem[Dedecker et~al\mbox{.}(2022)]%
        {dedecker2022s}
\bibfield{author}{\bibinfo{person}{Ruben Dedecker}, \bibinfo{person}{Wout Slabbinck}, \bibinfo{person}{Patrick Hochstenbach}, \bibinfo{person}{Pieter Colpaert}, {and} \bibinfo{person}{Ruben Verborgh}.} \bibinfo{year}{2022}\natexlab{}.
\newblock \showarticletitle{What’s in a Pod?--A knowledge graph interpretation for the Solid ecosystem}.
\newblock


\bibitem[Kahle(2015)]%
        {kahle2015locking}
\bibfield{author}{\bibinfo{person}{Brewster Kahle}.} \bibinfo{year}{2015}\natexlab{}.
\newblock \showarticletitle{{Locking the Web open: A call for a decentralized Web}}.
\newblock \bibinfo{journal}{\emph{Brewster Kahle’s Blog}} (\bibinfo{year}{2015}).
\newblock


\bibitem[Moaawad et~al\mbox{.}(2017)]%
        {moaawad2017fly}
\bibfield{author}{\bibinfo{person}{Mohamed~Ragab Moaawad}, \bibinfo{person}{Hoda~M O.~Mokhtar}, {and} \bibinfo{person}{Haytham~Tawfeek Al~Feel}.} \bibinfo{year}{2017}\natexlab{}.
\newblock \showarticletitle{On-the-fly academic linked data integration}. In \bibinfo{booktitle}{\emph{Proceedings of the International Conference on Compute and Data Analysis}}. \bibinfo{pages}{114--122}.
\newblock


\bibitem[Moawad et~al\mbox{.}(2019)]%
        {moawad2019minaret}
\bibfield{author}{\bibinfo{person}{Mohamed~R Moawad}, \bibinfo{person}{Mohamed Mohamed Maher Zenhom~Abdelrahman Maher}, \bibinfo{person}{Ahmed Awad}, {and} \bibinfo{person}{Sherif Sakr}.} \bibinfo{year}{2019}\natexlab{}.
\newblock \showarticletitle{Minaret: A recommendation framework for scientific reviewers}. In \bibinfo{booktitle}{\emph{the 22nd International Conference on Extending Database Technology (EDBT)}}.
\newblock


\bibitem[Ragab(2022)]%
        {ragab2022towards}
\bibfield{author}{\bibinfo{person}{Mohamed Ragab}.} \bibinfo{year}{2022}\natexlab{}.
\newblock \showarticletitle{Towards Prescriptive Analyses of Querying Large Knowledge Graphs}. In \bibinfo{booktitle}{\emph{European Conference on Advances in Databases and Information Systems}}. Springer, \bibinfo{pages}{639--647}.
\newblock


\bibitem[Ragab et~al\mbox{.}(2021)]%
        {ragab2021bench}
\bibfield{author}{\bibinfo{person}{Mohamed Ragab}, \bibinfo{person}{Feras~M Awaysheh}, {and} \bibinfo{person}{Riccardo Tommasini}.} \bibinfo{year}{2021}\natexlab{}.
\newblock \showarticletitle{Bench-ranking: a first step towards prescriptive performance analyses for big data frameworks}. In \bibinfo{booktitle}{\emph{2021 IEEE International Conference on Big Data (Big Data)}}. IEEE, \bibinfo{pages}{241--251}.
\newblock


\bibitem[Ragab et~al\mbox{.}(2023)]%
        {Ragab2023}
\bibfield{author}{\bibinfo{person}{Mohamed Ragab}, \bibinfo{person}{Yury Savateev}, \bibinfo{person}{Reza Moosaei}, \bibinfo{person}{Thanassis Tiropanis}, \bibinfo{person}{Alexandra Poulovassilis}, \bibinfo{person}{Adriane Chapman}, {and} \bibinfo{person}{George Roussos}.} \bibinfo{year}{2023}\natexlab{}.
\newblock \showarticletitle{ESPRESSO: A Framework for Empowering Search on Decentralized Web}. In \bibinfo{booktitle}{\emph{Web Information Systems Engineering -- WISE 2023}}, \bibfield{editor}{\bibinfo{person}{Feng Zhang}, \bibinfo{person}{Hua Wang}, \bibinfo{person}{Mahmoud Barhamgi}, \bibinfo{person}{Lu~Chen}, {and} \bibinfo{person}{Rui Zhou}} (Eds.). \bibinfo{publisher}{Springer Nature Singapore}, \bibinfo{address}{Singapore}, \bibinfo{pages}{360--375}.
\newblock
\showISBNx{978-981-99-7254-8}


\bibitem[Ragab et~al\mbox{.}(2020)]%
        {ragab2020towards}
\bibfield{author}{\bibinfo{person}{Mohamed Ragab}, \bibinfo{person}{Riccardo Tommasini}, \bibinfo{person}{Sadiq Eyvazov}, {and} \bibinfo{person}{Sherif Sakr}.} \bibinfo{year}{2020}\natexlab{}.
\newblock \showarticletitle{Towards making sense of Spark-SQL performance for processing vast distributed RDF datasets}. In \bibinfo{booktitle}{\emph{Proceedings of The International Workshop on Semantic Big Data}}. \bibinfo{pages}{1--6}.
\newblock


\bibitem[Sakr et~al\mbox{.}(2021)]%
        {sakr2021future}
\bibfield{author}{\bibinfo{person}{Sherif Sakr}, \bibinfo{person}{Angela Bonifati}, \bibinfo{person}{Hannes Voigt}, \bibinfo{person}{Alexandru Iosup}, \bibinfo{person}{Khaled Ammar}, \bibinfo{person}{Renzo Angles}, \bibinfo{person}{Walid Aref}, \bibinfo{person}{Marcelo Arenas}, \bibinfo{person}{Maciej Besta}, \bibinfo{person}{Peter~A Boncz}, {et~al\mbox{.}}} \bibinfo{year}{2021}\natexlab{}.
\newblock \showarticletitle{The future is big graphs: a community view on graph processing systems}.
\newblock \bibinfo{journal}{\emph{Commun. ACM}} \bibinfo{volume}{64}, \bibinfo{number}{9} (\bibinfo{year}{2021}), \bibinfo{pages}{62--71}.
\newblock


\bibitem[Sambra et~al\mbox{.}(2016a)]%
        {sambra2016building}
\bibfield{author}{\bibinfo{person}{Andrei Sambra}, \bibinfo{person}{Amy Guy}, \bibinfo{person}{Sarven Capadisli}, {and} \bibinfo{person}{Nicola Greco}.} \bibinfo{year}{2016}\natexlab{a}.
\newblock \showarticletitle{{Building decentralized applications for the social Web}}. In \bibinfo{booktitle}{\emph{Proceedings of the 25th International Conference Companion on World Wide Web}}. \bibinfo{pages}{1033--1034}.
\newblock


\bibitem[Sambra et~al\mbox{.}(2016b)]%
        {sambra2016solid}
\bibfield{author}{\bibinfo{person}{Andrei~Vlad Sambra}, \bibinfo{person}{Essam Mansour}, \bibinfo{person}{Sandro Hawke}, \bibinfo{person}{Maged Zereba}, \bibinfo{person}{Nicola Greco}, \bibinfo{person}{Abdurrahman Ghanem}, \bibinfo{person}{Dmitri Zagidulin}, \bibinfo{person}{Ashraf Aboulnaga}, {and} \bibinfo{person}{Tim Berners-Lee}.} \bibinfo{year}{2016}\natexlab{b}.
\newblock \showarticletitle{Solid: a platform for decentralized social applications based on linked data}.
\newblock \bibinfo{journal}{\emph{MIT CSAIL \& Qatar Computing Research Institute, Tech. Rep.}} (\bibinfo{year}{2016}).
\newblock


\bibitem[Taelman et~al\mbox{.}(2020)]%
        {Taelman2020}
\bibfield{author}{\bibinfo{person}{Ruben Taelman}, \bibinfo{person}{Simon Steyskal}, {and} \bibinfo{person}{Sabrina Kirrane}.} \bibinfo{year}{2020}\natexlab{}.
\newblock \showarticletitle{Towards Querying in Decentralized Environments with Privacy-Preserving Aggregation}. In \bibinfo{booktitle}{\emph{Joint Proceedings of Workshops AI4LEGAL2020, NLIWOD, PROFILES 2020, QuWeDa 2020 and SEMIFORM2020 Colocated with the 19th International Semantic Web Conference (ISWC 2020)}}, \bibfield{editor}{\bibinfo{person}{{Manolis Koubarakis, Harith Alani, Grigoris Antoniou, Kalina Bontcheva, John Breslin, Diego Collarana, Elena Demidova, Stefan Dietze, Simon Gottschalk, Guido Governatori, Aidan Hogan, Freddy Lecue, Elena Montiel Ponsoda, Axel-Cyrille Ngonga Ngomo, Sofia Pinto, Muhammad Saleem, Raphael Troncy, Eleni Tsalapati, Ricardo Usbeck, Ruben Verborgh}}} (Ed.). \bibinfo{pages}{135 -- 148}.
\newblock


\bibitem[Tiropanis et~al\mbox{.}(2021)]%
        {tiropanis2021search}
\bibfield{author}{\bibinfo{person}{Thanassis Tiropanis}, \bibinfo{person}{Alexandra Poulovassilis}, \bibinfo{person}{Age Chapman}, {and} \bibinfo{person}{George Roussos}.} \bibinfo{year}{2021}\natexlab{}.
\newblock \showarticletitle{Search in a Redecentralised Web}. In \bibinfo{booktitle}{\emph{Computer Science Conference Proceedings: 12th International Conference on Internet Engineering; Web Services (InWeS 2021)}}.
\newblock


\bibitem[Zhao et~al\mbox{.}(2024)]%
        {zhao2024llm}
\bibfield{author}{\bibinfo{person}{Jujia Zhao}, \bibinfo{person}{Wenjie Wang}, \bibinfo{person}{Chen Xu}, \bibinfo{person}{Zhaochun Ren}, \bibinfo{person}{See-Kiong Ng}, {and} \bibinfo{person}{Tat-Seng Chua}.} \bibinfo{year}{2024}\natexlab{}.
\newblock \showarticletitle{LLM-based Federated Recommendation}.
\newblock \bibinfo{journal}{\emph{arXiv preprint arXiv:2402.09959}} (\bibinfo{year}{2024}).
\newblock


\end{thebibliography}

\end{document}